\documentclass[10pt,letterpaper]{article}
\usepackage{multirow}
\usepackage{graphicx}
\usepackage{amsmath}
\usepackage{cases}
\usepackage{amssymb}
\usepackage{enumerate}
\usepackage[compress]{cite}

\begin{document}

\title{Behaviour of light transmission channels in random media with inhomogeneous disorder}

\author{Yuchen Xu, Hao Zhang$^{*}$,Yujun Lin and Heyuan Zhu$^\dag$\\ \textit{\small Shanghai Ultra-precision Optical Manufacturing Engineering Center,} \\ \textit{\small Department of Optical Science and Engineering,} \\ \textit{\small Fudan University, Shanghai 200433, China}\\ \small $^*$zhangh@fudan.edu.cn\\ \small $^\dag$hyzhu@fudan.edu.cn}

\date{}

\maketitle

\begin{abstract}
We present a numerical study on the light transport properties and statistics of transmission channels in random media with inhomogeneous disorder. For the case of longitudinal inhomogeneity of disorder we find that the statistics of the transmission channels is independent of the inhomogeneity and the system can be equivalent to a counterpart with homogeneous disorder strength, both of which have the same statistical distribution of the transmission channels. However, for the case of transverse inhomogeneity of disorder, such equivalence does not exist, moreover, the transmission eigenvalues are pushed to the two ends of the distribution and the distribution of the total transmission is broadened since the spatial structure gives rise to larger and smaller transmitted incident channels.
\end{abstract}

\section{Introduction}
Coherent transport of waves in random media at the the mesoscopic scale showing extraordinary characteristics has attracted much attention in recent decades. Wave interference during the multiple scattering leads to many amazing physical phenomena, such as Anderson localization, enhanced backscattering and universal conductance fluctuations\cite{Anderson1958,Albada1985,Wolf1985,Lee1985,Scheffold1998}. In the theoretical framework of quantum transport, an incident wave (outgoing wave) can be decomposed into several transport channels, which correspond to the quantized directions in which the wave enters (exits) the random medium. 

The transmission behaviour of the channels is governed by the $N \times N$ field transmission matrix $t$, where $N$ is the number of the channels. With $t$ one can obtain the transmitted intensity, the total transmission for different incident channels as well as the transmittance. From the statistics of these quantities one can extract rich information about the transmitted wave\cite{Rossum1999}. Eigenvalues $\{\tau_{n}\}$ of the Hermitian matrix $t^{\dagger}t$ together with the corresponding eigenchannels can also be extracted from $t$\cite{Dorokhov1982,Dorokhov1984,Mello1988}, and researches on the so-called ``open channels'' with $\tau_{n} \simeq 1$ has realized focusing and imaging of light through turbid media\cite{Vellekoop2007,Vellekoop2008,Popoff2010,Choi2012,Kim2012,Kim2013}. 

The transmission matrix $t$ is determined by the configuration of disorder when the dimensions of the sample are fixed, and previous researches mostly focused on homogeneous disorder. However, inhomogeneous disorder exists widely in natural and artificial materials. The inhomogeneity may results from multilayer configurations or inhomogeneous doping, which are common in real experiments and can not be eliminated by ensemble averaging. 

The additional degree of freedom introduced by the inhomogeneity of disorder will cause difficulties for theoretical investigations on transport properties and the behaviour of channels in such random media. Thus two fundamental questions are proposed. The first question is that, is there any equivalent treating method, which is analogous to the effective medium theory, and can be applied to deal with the inhomogeneity? And the subsequent question is that, if there is no such method, then how can we consider the influence of such inhomogeneity? 

In this work, we investigate the influence of inhomogeneous disorder by numerical simulations and try to provide basic comprehension for the questions proposed above. 

\section{Methods and Configurations}
To study the light transport properties and the behaviour of transmission channels in random media, we consider a 2D disordered waveguide with two semi-infinite free waveguides attached to its both sides. The transverse boundaries of the entire system are perfectly reflective. A monochromatic scalar wave $\psi(x,y)e^{-i\omega t}$ propagates along the longitudinal direction from left to right in the system and the propagation is governed by the Helmholtz equation 
\begin{equation}
\left[ \nabla^2 + k^2 \varepsilon(x,y) \right] \psi(x,y) = 0, \label{helmholtz}
\end{equation}
where $k=\omega/c$ is the wave number, $c$ is the wave speed in vacuum. $\varepsilon(x,y)=1+\delta\varepsilon(x,y)$ is the relative dielectric constant with a randomly fluctuation $\delta\varepsilon(x,y)$, which is uniformly distributed between $[-\sigma,\sigma]$ in the scattering region and equals to zero outside the scattering region. 

\begin{figure}[ht]
\centering
\includegraphics[scale=0.35]{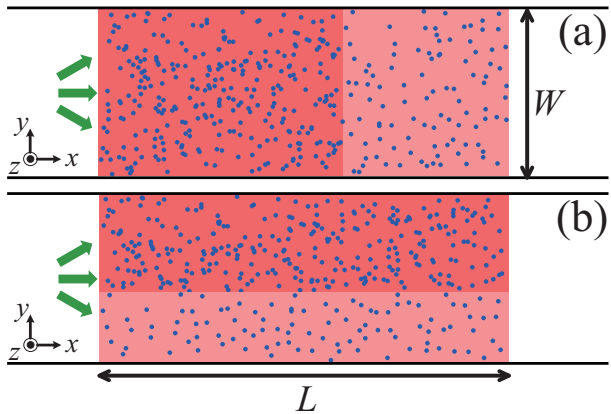}
\caption{Schematics of 2D disordered waveguides with (a) longitudinal (b) transverse inhomogeneity of disorder. }
\label{Fig:multi}
\end{figure}

The element $t_{ba}$ of the transmission matrix $t$ of the entire system, which represents the complex field transmission amplitude from the incoming channel $a$ to the outgoing channel $b$, can be calculated with the Fisher-Lee relation\cite{Fisher1981} 
\begin{equation}
t_{ba}=\sqrt{v_{b}v_{a}}\int_{0}^{W}\mathrm{d}y \int_{0}^{W}\mathrm{d}y' \chi_{b}^{*}(y)G^{\mathrm{r}}(L,y;0,y')\chi_{a}(y'), \label{telement}
\end{equation}
where $v_{n}$ is the group velocity at the incident wavelength of the $n^{\mathrm{th}}$ channel in the free waveguide, $\chi_{n}(y)$ is the corresponding transverse wave function, which takes the standing wave form due to perfectly reflective boundaries. $G^{\mathrm{r}}(L,y;0,y')$ is the retarded Green's function from the source point $(0,y')$ to the probe point $(L,y)$, which can be calculated using the recursive Green's function(RGF) method\cite{MacKinnon1985,Baranger1991}. In this method, the wave equation is discretized using a 2D tight-binding model on a square lattice with spacing constant $d$. 

The transmitted intensity $T_{ba}$, the total transmission $T_{a}$ and the transmittance $T$ can be calculated directly from the elements of $t$ as 
\begin{equation}
T_{ba} = \left| t_{ba} \right|^{2},\hspace{1cm}T_{a} = \sum_{b=1}^{N}\left|t_{ba}\right|^{2},\hspace{1cm}T = \sum_{a=1}^{N}\sum_{b=1}^{N} \left|t_{ba}\right|^{2}. 
\end{equation}
$T_{ba}$ determines the transmitted speckle pattern which results from the incident channel $a$ and the output channel $b$, $T_{a}$ corresponds to the brightness of the speckle pattern induced by incident channel $a$, and $T$ is the classical counterpart of the electronic dimensionless conductance $g$. 

The singular value decomposition of the field transmission matrix $t=U \cdot \sqrt{\mathrm{diag}\{\tau_{n}\}} \cdot V^{\dagger}$ gives the $N$ transmission eigenvalues $\{\tau_{n}\}$ of $t^{\dagger}t$, where $U$ and $V$ are unitary matrices which map the eigenchannels in the disordered region to the output channels and incident channels, respectively\cite{Mello1988}. The transmittance $T$ can also be obtained by summing over all the transmission eigenvalues, i.e., $T=\sum_{n=1}^{N}\tau_{n}$. For simplicity, the ensemble average of the transmittance $\left<T\right>$ will be denoted by $g$ without ambiguity. 

An important statistical description of the transport property of a disordered waveguide is the distribution of the transmission eigenvalues $\rho(\tau)$, which is defined as 
\begin{equation}\label{dte}
\rho(\tau)=\frac{1}{N}\left< \sum_{n=1}^{N}\delta(\tau-\tau_{n})\right>. 
\end{equation}
For disordered samples with homogeneous disorder, $\rho(\tau)$ is bimodal in the diffusive regime, as follows 
\begin{equation}\label{bimodal}
\rho(\tau)=\rho_{0}(\tau) \equiv \frac{\overline{\tau}}{2}\frac{1}{\tau \sqrt{1-\tau}}, 
\end{equation}
where $\overline{\tau}=g/N$ is the averaged transmittance. The mode peak at $\tau \simeq 1$ corresponds to the ``open channels'' and the peak at $\tau \simeq 0$ corresponds to the ``closed channels''. 

To analyse the influence of inhomogeneous disorder by simulation, we considered two standard configurations of inhomogeneous disorder, as shown in Fig. \ref{Fig:multi}. In the first configuration (Fig. \ref{Fig:multi}(a)) the disordered region is divided into two layers arranged in the longitudinal direction, with the length of the left layer being $L_{1}=fL$ and the magnitude of $\delta\varepsilon(x,y)$ equal to $\sigma_{1}$ in the left layer and $\sigma_{2}$ in the right layer, respectively. Analogously, in the second configuration (Fig. \ref{Fig:multi}(b)) the disordered region is divided into two layers arranged in the transverse direction, with the width of the upper layer being $W_{1}=fW$ and $\sigma_{1}$ and $\sigma_{2}$ equal to the fluctuation magnitudes of the upper and the lower layers, respectively. In the following simulations the wave vector is $k=1.5/d$, and all the lengths are scaled in units of $d$. 

\section{Numerical Results and Discussions}
\subsection{Longitudinal inhomogeneity of disorder}
The conductance $g$ is determined by the scaling parameter $L/\xi$ for random media with homogeneous disorder, where the localization length $\xi$ is related to the mean free path $l$ by the Thouless relation $\xi \simeq Nl$ in the Q1D limit, i.e., $l \gg W$\cite{Dorokhov1982}. The mean free path is inversely proportional to the disorder strength $\sigma$, i.e., $l=l_{0}\sigma^{-2}$, with $l_{0}$ being the mean free path for $\sigma=1$\cite{xu2015}. Thus when the size of the sample is fixed, $g$ is the function of the single parameter $\sigma$. While for samples with inhomogeneous disorder, theoretically $g$ should depend on the parameters $\sigma_{1}$, $\sigma_{2}$ and $f$. 

Generally, two disordered samples which support the same number of eigenchannels can be treated as equivalent in respect of light transport, when they give same conductance $g$ and eigenvalue density distribution $\rho(\tau)$ under the same incident light condition. For samples with homogeneous disorder, the same $g$ just intrinsically means the same $\rho(\tau)$, while for samples with inhomogeneous disorder, intuitively, $g$ can not exclusively determine $\rho(\tau)$, considering that the disorder configuration probably influences the transport behaviours of different eigenchannels. 

We introduce an effective disorder strength $\overline{\sigma}$ to describe the disorder strength of random media with longitudinal disorder inhomogeneity, which is defined as 
\begin{equation}
\overline{\sigma}^{2}=f\sigma_{1}^{2}+(1-f)\sigma_{2}^{2}. \label{effdis}
\end{equation}

\begin{figure}[ht]
\centering
\includegraphics[scale=0.35]{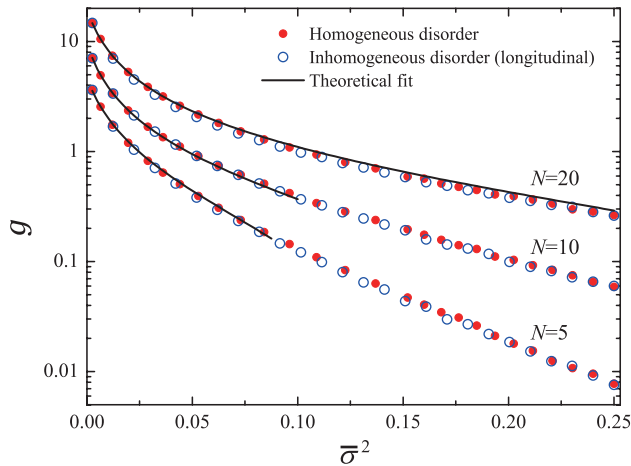}
\caption{Classical conductance $g$ calculated as a function of the square of the effective homogeneous disorder strength for samples with homogeneous (red filled circles) and longitudinally inhomogeneous (blue empty circles) disorder, for 3 channel numbers $N=5, 10, 20$ (from bottom to top). The solid lines are theoretical fits according to Eq. \eqref{gexpansion}.}
\label{Fig:muf}
\end{figure}

To compare $\overline{\sigma}$ with $\sigma$ (samples with homogeneous disorder strength $\sigma$), the conductance $g$ is calculated for both cases of homogeneous (red filled circles) and inhomogeneous (blue empty circles) disorder, for three different channel numbers $N=5,10,20$, as shown in Fig. \ref{Fig:muf}. The length $L$ of samples is fixed at $400d$. The disorder strength under consideration is $\sigma_{1}=0.5$ and $\sigma_{2}=0.05$. The ensemble averages are performed over 20000, 10000 and 5000 random realizations for the three channel numbers, respectively. 

As we know, in the delocalized regime, $s \ll N$, where $s=L/l$, the calculated $g$ can be fitted in the Q1D limit by the perturbative expansion\cite{Mirlin2000,Payne2010}
\begin{equation}
g = g_{0} - \frac{1}{3} + \frac{1}{45g_{0}} + \frac{2}{945g_{0}^{2}} + \mathcal{O}\left( \frac{1}{g_{0}^3} \right), \label{gexpansion}
\end{equation}
where $g_0 = N(1+s)^{-1}$ is the leading term of $g$ which takes into account the so-called extrapolation length $z_0=l/2$ induced by the internal reflection at the open boundaries of the disordered region\cite{Rossum1999}. 

It is shown in Fig. \ref{Fig:muf} that, by taking $l_0$ as the only fitting parameter with the values of $l_0=3.5d, 2.9d, 3.1d$ for $N=5,10,20$ respectively, the theoretic values (solid lines) obtained by Eq. \eqref{gexpansion} fit well to the numerical results in the delocalized regime. When the effective disorder strength  $\overline{\sigma}$ of the Q1D random systems is large enough, the corresponding transport enters the strongly localized regime and thus the relation described by Eq. \eqref{gexpansion} is no longer valid. Here for the Q1D samples with $N=5$ and $N=10$, the critical values of $\overline{\sigma}$ are near 0.27 and 0.32, respectively. For the sample with $N=20$, the transport under considerations shown in  Fig. \ref{Fig:muf} is not beyond the weak-localization limit, however the random system gradually changes from the Q1D limit to the general 2D configuration around $\overline{\sigma}^2=0.1$ due to $l \simeq W$, thus the theoretic curve also slowly deviates from the numerical result when $\overline{\sigma}^2 > 0.1$. 

Comparison between the theoretical fits and the numerical results obviously shows that the classical conductance of the sample with inhomogeneous disorder along the longitudinal direction is the same as that of the one with homogeneous disorder with the introduction of the \textit{effective disorder strength} $\overline{\sigma}$. 

\begin{figure}[ht]
\centering
\includegraphics[scale=0.35]{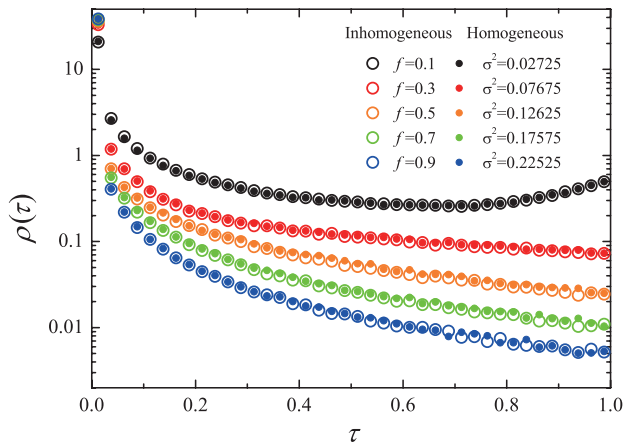}
\caption{Distributions of eigenvalue density for samples with longitudinal inhomogeneity of disorder and homogeneous disorder calculated for $f=0.1,0.3,0.5,0.7,0.9$ (empty circles, from top to bottom) in the former case and corresponding values of $\sigma^{2}$ obtained from Eq. \eqref{effdis} in the latter case (filled circles, from top to bottom).}
\label{Fig:dtel}
\end{figure}

To verify the validity of the mentioned equivalence, it is necessary to investigate the distribution of the transmission eigenvalue density $\rho(\tau)$. The calculated transmission eigenvalue densities for the case of $N=10$ as in Fig. \ref{Fig:muf} are shown in Fig. \ref{Fig:dtel}, where empty circles correspond to the inhomogeneous disorder and filled circles correspond to effective homogeneous disorder, and the fraction factor $f$ together with the corresponding effective disorder $\overline{\sigma}^2$ calculated with Eq. \eqref{effdis} are listed as legends in Fig. \ref{Fig:dtel}. It is shown that the eigenvalue density of the inhomogeneous-disorder case is in good consistence with the one of the effective homogeneous-disorder case, which indicates that the inhomogeneity does not modify the transmission eigenchannels and thus reveals that longitudinally inhomogeneous disorder is definitely equivalent to homogeneous disorder considering the identical statistics of the transmission channels. 

Futhermore, the universal equivalence can be understood in the framework of the Dorokhov-Mello-Pereyra-Kumar (DMPK) equation\cite{Dorokhov1984,Mello1988,Beenakker1997} , as follows, 
\begin{equation}
l\frac{\partial p(\{\lambda_{n}\};L)}{\partial L}=\frac{2}{N+1}\frac{1}{J(\{\lambda_{n}\})}\sum_{n=1}^{N}\frac{\partial}{\partial \lambda_{n}}\left[ \lambda_{n}(1+\lambda_{n})J(\{\lambda_{n}\})\frac{\partial p(\{\lambda_{n}\};L)}{\partial \lambda_{n}}\right], \label{dmpk}
\end{equation}
where $p(\{\lambda_{n}\};L)$ is the joint probability density of the $N$ random variables $\{\lambda_{n}\}$ which are the parametrization variables $\lambda_{n}$ representing the ratio of the reflection to transmission probabilities of each eigenchannel and satisfying 
\begin{equation}
\lambda_{n}=\frac{1-\tau_{n}}{\tau_{n}}, \label{parametrization}
\end{equation}
and 
\begin{equation}
J(\lambda)=\prod_{n<m}\left| \lambda_{n}-\lambda_{m} \right|
\end{equation}
is the Jocobian which results from the transform from the transfer matrix to diagonal matrix whose elements are the transmission eigenvalues. This equation describes the evolution of $p(\{\lambda_{n}\};L)$ with the increasing sample length $L$ by means of the transfer matrix method (TMM), with which one can obtain $\rho(\tau)$ immediately and $g$ by ensemble-averaging. 

Since Eq. \eqref{dmpk} is initially developed to solve the transport problem of random media with homogeneous disorder, therefore the mean free path $l$ is invariant along the longitudinal direction. Here our numerical results indicate that it can be applied to the case of longitudinally inhomogeneous disorder. 

By taking $s=L/l$ as the independent variable but not just an abbreviation which contributes to the evolution of $p(\{\lambda_{n}\};s)$, Eq. \eqref{dmpk} can be integrated on both sides with the initial condition $p(\{\lambda_{n}\};s=0)=\delta(\{\lambda_{n}\})$ to give Eq. \eqref{gexpansion}, no matter $l$ varies along the longitudinal direction or not. By this way the effective disorder strength can be generalized to the case of continuously varying disorder along the longitudinal direction, i.e., $\sigma(x)$, with the integral form 
\begin{equation}
\overline{\sigma}^{2}=\frac{1}{L}\int_{0}^{L} \left[\sigma(x)\right]^{2}\mathrm{d}x. \label{generaldis}
\end{equation}

In fact, Eq. \eqref{dmpk} is derived based on the hypothesis that the light transport is isotropic, which implies that the flux incident in a given channel is scattered into any channel with the same probability. This hypothesis keeps valid as long as the disorder is homogeneous in the transverse direction, despite the inhomogeneity in the longitudinal direction. Moreover, since the reflection probability per unit sample length is equal to the inverse of the mean free path $l$\cite{Mello1991}, and thus is proportional to $[\sigma(x)]^{2}$\cite{xu2015}, then by taking the scattering effect of all length units into account one can obtain Eq. \eqref{generaldis}, which is actually a reasonable extrapolation of the longitudinally position-dependent mean free path. 

\subsection{Transverse inhomogeneity of disorder}
When the two layers with different disorder strength are arranged along the transverse direction, as shown in Fig. \ref{Fig:multi}(b), the situation is more complicated and interesting. First it is important to check whether there as well exists any strict equivalence between the cases of homogeneous and inhomogeneous disorder by comparing the respective distributions of the transmission eigenvalue density $\rho(\tau)$. 

The random samples size $L$ and $W$ are fixed at $400d$ and $50d$, and the disorder strengths of the upper and lower layer are $\sigma_{1}=0.5$ and $\sigma_2=0.05$, respectively. Four samples are under consideration and labelled as $A1-A4$, with the width fraction of the upper layer $f=0.2,0.5,0.7,0.9$ in sequence. 

\begin{figure}[ht]
\centering
\includegraphics[scale=0.35]{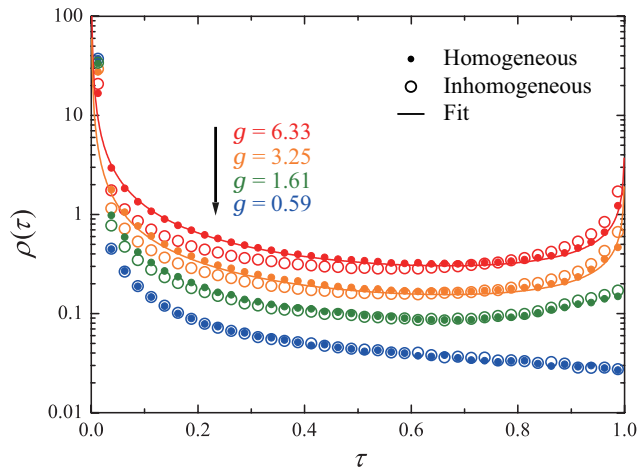}
\caption{Distributions of eigenvalue density for samples with transversely inhomogeneous disorder and homogeneous disorder calculated for $f=0.2,0.5,0.7,0.9$ (empty circles, from top to bottom) in the former case and corresponding values of $\sigma^{2}$ obtained for same $g$'s in the latter case (filled circles, from top to bottom).}
\label{Fig:dtel2}
\end{figure}

The calculated conductance $g$ for the four samples are 6.33, 3.25, 1.61 and 0.59, respectively. To verify the above-mentioned equivalence, we calculated   conductance $g$ of a series of random samples with homogeneous disorder strength and same sample sizes, and finally obtained four random samples with identical respective conductance $g$ of $A1-A4$. The disorder strengths for the obtained four samples are approximately $\sigma=0.152,0.221,0.305,0.454$, respectively, and we labeled these four samples as $B1-B4$. The calculated distributions of transmission eigenvalue density $\rho(\tau)$ for samples $A1-A4$ and $B1-B4$ are plotted in circles in Fig. \ref{Fig:dtel2}. 

Since the solid lines in Fig. \ref{Fig:dtel2} calculated theoretically using Eq. \eqref{bimodal} with the corresponding $g$ fit well to the numerical results for homogeneous-disorder samples B1 and B2, therefore light transport in these two samples is in the diffusive regime. However, as a comparison, for the corresponding inhomogeneous-disorder samples A1 and A2, the distributions of eigenvalue density are obviously different from the theoretical values, especially for large and very small (close to 0) eigenvalues. Similar behaviour can be found by comparing $\rho(\tau)$ of samples A3 and B3.

\begin{figure}[ht]
\centering
\includegraphics[scale=0.4]{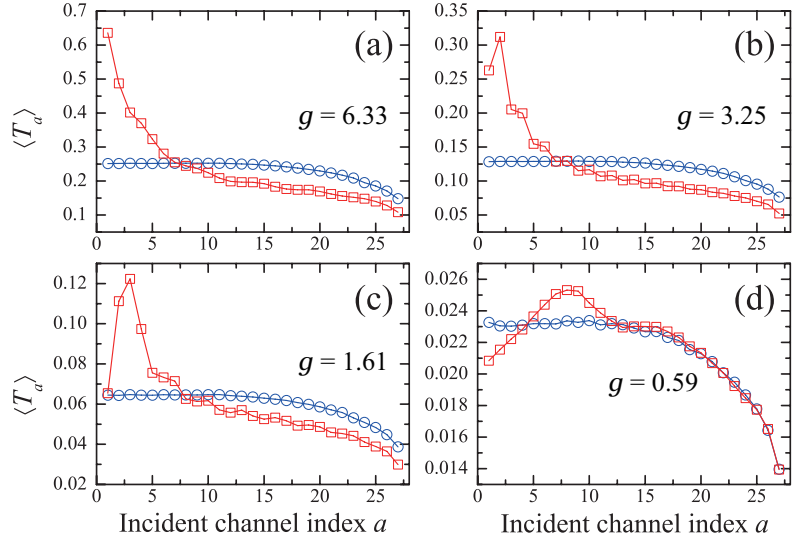}
\caption{The ensemble-averaged total transmission $\left<T_{a}\right>$ of different incident channels for samples with (a) $g=6.33$, (b) $g=3.25$, (c) $g=1.61$ and (d) $g=0.59$. The blue connected circles represent the samples with homogeneous disorder and the connected squares represent the samples with transversely inhomogeneous disorder. }
\label{Fig:ttt}
\end{figure}

However, for samples A4 and B4 with the smallest conductance $g$ among these random samples, the difference in $\rho(\tau)$ almost disappears. Futher investigation on the total transmission for different incident channel $\left<T_{a}\right>$ of samples A4 and B4, as shown in Fig.\ref{Fig:ttt}(d), reveals obviously different transport properties between them.   

As a result, by the comparison among transmission eigenvalue distributions $\rho(\tau)$, it is found that when the width fraction $f$ is non-trivial, even though two random samples with homogeneous and inhomogeneous disorder give identical conductance $g$, the statistics of eigenchannels for them are probably different. Thus it can be concluded that random samples with transverse inhomogeneity of disorder definitely can not be equivalent to those with homogeneous disorder, which implies that the scaling theory is no longer valid under such situation. 

In order to shed light on the anomalous transport properties of random samples with transversely inhomogeneous disoder, here we consider an extreme case in which the disorder strength of the lower layer equals zero, i.e., the lower layer is scattering free. Light in incident channels with small incident angles may travel through the lower layer without experiencing any scattering, thus the status of different incident channels are totally different. 

In this sense, it is more instructive to study the statistics of the total transmission  $T_{a}$ for different incident channels $a$. The incident channel index $a$ represents the transverse mode of the corresponding incident channel which corresponds to the incident angle $\theta_{a} = \cos ^{-1}(k_{a}/k)$ ($k_{a}$ is the longitudinal wave vector). 

The dependence of the averaged total transmission $\left<T_{a}\right>$ on the channel index $a$ is shown in Fig.~\ref{Fig:ttt}. Here the average is only taken over different random configurations, and thus $\left<T_{a}\right>$ represents the ratio of light that transmitted through the scattering region on average for a certain incident angle $\theta_{a}$. For the case of homogeneous disorder (samples $B1-B4$), $\left<T_{a}\right>$ hardly depends on the incident channel index $a$ in a wide range ($a\lesssim 15$ herein), which means that in random samples with homogeneous disorder, the probability of light transport through relatively high transmission channels ($a<15$ or $\theta<0.22\pi$ for homegeneous disorder samples $B1-B4$) is nearly equal, regardless of the disorder strength of random samples.

While for the case of transversely inhomogeneous disorder samples (i.e. $A1-A4$), $\left<T_{a}\right>$ considerably depends on the incident channel $a$, which shows a relatively larger variation, and as well as a transmission peak corresponding to the most transmitted incident channel, as shown in Fig.~\ref{Fig:ttt}. 

When the fraction of the weakly scattering layer is large enough, e.g. $f=0.2$ shown in Fig.~\ref{Fig:ttt}(a), the highest transmission takes place at the incident channel with the smallest incident angle, i.e. $a=1$. When the fraction of the strongly scattering layer $f$ increases, the index of the most transmitted channel increases ($a=2,3,8$ shown in Fig.~\ref{Fig:ttt}(b-d)), which means that the incident channel with the smallest incident angle is no longer the most transmitted channel, but replaced by another channel with a larger incident angle.

\begin{figure}[ht]
\centering
\includegraphics[scale=0.4]{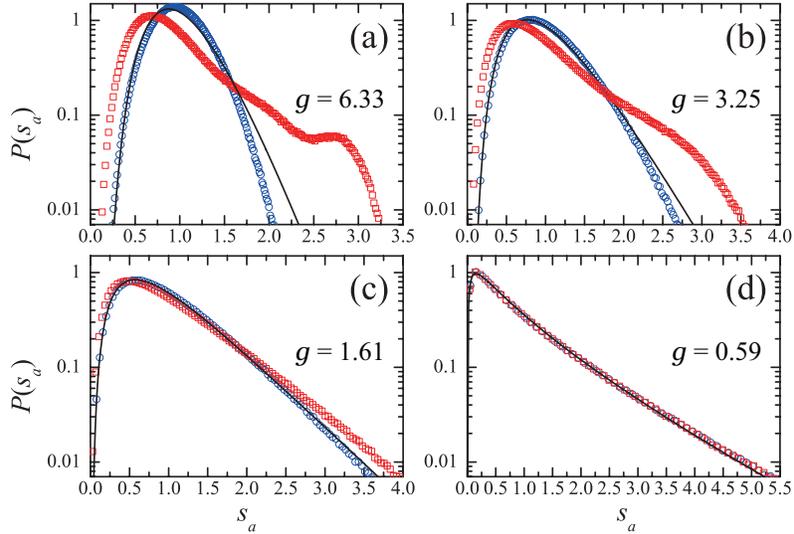}
\caption{Distributions of normalized total transmission $s_{a}$ for samples with (a) $g=6.33$, (b) $g=3.25$, (c) $g=1.61$ and (d) $g=0.59$. The blue circles represent the samples with homogeneous disorder, the squares represent the samples with transversely inhomogeneous disorder and the solid lines are the predictions of Eq.~\eqref{psaphi}. }
\label{Fig:psa}
\end{figure}

The distribution of the total transmission should also be modulated by the spatial structure of the transversely inhomogeneous disorder. The calculated probability densities of the normalized total transmission $s_{a}=T_{a}/\overline{\tau}$ (here the subscript ``$a$'' is maintained to distinguish with the transmittance $T$ and takes all the incident channels into account), i.e., $P(s_{a})$, are shown in Fig. \ref{Fig:psa} for different conductance $g$. 

Theoretical prediction for $P(s_{a})$ depends on the single parameter $g$ as follows\cite{Nieuwenhuizen1995},
\begin{subequations}\label{psaphi}
\begin{align}
P(s_{a}) & =\int_{-i\infty}^{i\infty}\frac{\mathrm{d}x}{2\pi i}\exp \left[ xs_{a}-\Phi(x)\right], \label{psa}\\
\Phi(x) & =g \mathrm{ln}^{2}\left( \sqrt{1+x/g}+\sqrt{x/g} \right). \label{phi}
\end{align}
\end{subequations}

The theoretical distributions of $s_{a}$ of random samples with homogeneous disorder ($B_1-B_4$) are obtained and shown as black lines in Fig. \ref{Fig:psa}, which match well with the numerical distributions shown as blue circles in Fig. \ref{Fig:psa}. For the case of an arbitrary conductance $g$, $P(s_{a})$ shows an exponentially decaying tail. When the conductance $g$ is large enough, e.g., $g=6.33$ in Fig.~\ref{Fig:psa}(a), the shape of the probability $P(s_{a})$ is Gaussian-like in the vicinity of $s_{a} \sim 1$, and becomes log-normal when $s_{a} \ll 1$. When $g$ decreases, the transmission peak deviates from $s_{a} \sim 1$ and locates at some value of $s_{a}<1$, and the Gaussian shape around the peak gradually vanishes. 

As shown in Fig.~\ref{Fig:psa}, the distributions of $s_{a}$ of the samples with inhomogeneous disorder are broader than those of the corresponding samples with homogeneous disorder (except for A4 and B4), even though they have the same conductance $g$. This phenomenon means that, when the fraction of the weakly scattering layer is large enough, light incident from different directions to occupy a larger range of transmittance. 

For sample A1 with $f=0.2$, there is another peak of $P(s_{a})$ near $s_{a}=2.75$, which corresponds to the average transmission peak at $a=1$ in Fig.~\ref{Fig:ttt}(a). Similarly, $s_{a}\simeq 2.75$ in Fig.\ref{Fig:psa}(b) corresponds to the peak at $a=2$ in Fig.~\ref{Fig:ttt}(b). This means that some of the channels are hardly influenced by the upper layer with strong disorder. When $f$ increases, the extra peak gradually vanishes and the distribution of $s_{a}$ approaches that of the sample with homogeneous disorder. When $f$ is large enough, the effect of inhomogeneity nearly vanishes. 

Note that though the influence of the spatial structure is not revealed by considering the distribution of total transmission as shown in Fig.~\ref{Fig:psa}(d), it still considerably affects the statistics of the light transmission which has been shown in Fig.~\ref{Fig:ttt}(d). 

\section{Conclusions}
We have carried out a detailed numerical investigation on how the inhomogeneity of disorder influence the light transport properties and the statistics of transmission channels in 2D disordered waveguides. For waveguides with longitudinal inhomogeneity of disorder, transmission channels are not modified and the transport of light can be equivalent to that in waveguides with effective homogeneous disorder. However, for waveguides with transverse inhomogeneity of disorder, the statistics of the transmission channels are considerably modified and the light transport reveals hybrid behaviours of different regimes, which leads to the additional repulsion of large and small transmission eigenvalues and broadening of the distributions of the total transmission. The results in the present paper may promote both the theoretical and experimental investigations on light transport in more extensive disordered materials. 

\section*{Acknowledgements}
This work is supported by the National Natural Science Foundation of China under Grant No. 11374063, and 973 Program(No. 2013CAB01505).

\end{document}